\documentclass[twocolumn]{aastex62}

\newcommand\latex{La\TeX}

\received{January 8, 2019}
\revised{February 27, 2019}
\accepted{}
\submitjournal{ApJ}

\shorttitle{Two-Molecule Combination Band}
\shortauthors{Tegler et al.}

\begin{document}

\title{A New Two-Molecule Combination Band as Diagnostic of Carbon Monoxide Diluted in Nitrogen Ice On Triton}

\correspondingauthor{S.C. Tegler}
\email{stephen.tegler@nau.edu}

\author{S.C. Tegler}
\affiliation{Northern Arizona University \\
Department of Physics and Astronomy \\
Flagstaff, AZ, 86011, USA}

\author{T.D. Stufflebeam}
\affiliation{Northern Arizona University \\
Department of Physics and Astronomy \\
Flagstaff, AZ, 86011, USA}

\author{W.M. Grundy}
\affiliation{Lowell Observatory \\
Mars Hill, Flagstaff, AZ, 86001, USA}
\affiliation{Northern Arizona University \\
Department of Physics and Astronomy \\
Flagstaff, AZ, 86011, USA}

\author{J.Hanley}
\affiliation{Lowell Observatory \\
Mars Hill, Flagstaff, AZ, 86001, USA}
\affiliation{Northern Arizona University \\
Department of Physics and Astronomy \\
Flagstaff, AZ, 86011, USA}

\author{S. Dustrud}
\affiliation{Northern Arizona University \\
Department of Chemistry and Biochemistry \\
Center for Materials Interfaces in Research and Applications\\
Flagstaff, AZ, 86011, USA}

\author{G.E. Lindberg}
\affiliation{Northern Arizona University \\
Department of Chemistry and Biochemistry \\
Center for Materials Interfaces in Research and Applications\\
Flagstaff, AZ, 86011, USA}

\author{A. Engle}
\affiliation{Northern Arizona University \\
Department of Physics and Astronomy \\
Flagstaff, AZ, 86011, USA}

\author{T. R. Dillingham}
\affiliation{Northern Arizona University \\
Department of Physics and Astronomy \\
Flagstaff, AZ, 86011, USA}

\author{D. Matthew}
\affiliation{Northern Arizona University \\
Department of Chemistry and Biochemistry \\
Flagstaff, AZ, 86011, USA}

\author{D. Trilling}
\affiliation{Northern Arizona University \\
Department of Physics and Astronomy \\
Flagstaff, AZ, 86011, USA}

\author{H. Roe}
\affiliation{Association of Universities for Research in Astronomy \\
Gemini Observatory, Chile}

\author{J.Llama}
\affiliation{Lowell Observatory \\
Mars Hill, Flagstaff, AZ, 86001, USA}

\author{G. Mace}
\affiliation{University of Texas \\
Department of Astronomy \\
Austin, TX, 78712, USA}

\author{E. Quirico}
\affiliation{Universite Grenoble Alpes \\
Grenoble, France}



\begin{abstract}

A combination band due to a mechanism whereby a photon excites two or more vibrational modes ({\it e.g.} a bend and a stretch) of an individual molecule  is commonly seen in laboratory and astronomical spectroscopy. Here, we present evidence of  a much less commonly seen combination band $-$ one where a photon simultaneously excites two adjacent molecules in an ice. In particular, we present near-infrared spectra of laboratory CO/N$_2$ ice samples where we identify a band at 4467.5 cm$^{-1}$ (2.239 $\mu$m) that results from single photons exciting adjacent pairs of CO and N$_2$ molecules.  We also present a near-infrared spectrum of Neptune's largest satellite Triton taken with the Gemini-South 8.1 meter telescope and the Immersion Grating Infrared Spectrograph (IGRINS) that shows this 4467.5 cm$^{-1}$ (2.239 $\mu$m) CO-N$_2$ combination band. The existence of the band in a spectrum of Triton indicates that CO and N$_2$ molecules are intimately mixed in the ice rather than existing as separate regions of pure CO and pure N$_2$ deposits. Our finding is important because CO and N$_2$ are the most volatile species on Triton and so dominate seasonal volatile transport across its surface. Our result will place constraints on the interaction between the surface and atmosphere of Triton. 

\end{abstract}


\keywords{methods: laboratory --- methods: observational --- planets and satellites: surfaces --- techniques: spectroscopic}


\section{Introduction} 
Carbon monoxide (CO) ice and nitrogen (N$_2$) ice have been detected in ground-based telescope spectra of Triton \citep{cru84,cru93} and Pluto \citep{owe93}. An unanswered question is whether CO molecules are diluted in N$_2$ ice or whether CO and N$_2$ exist as separate and pure ices. Answering this question is important because CO and N$_2$  are the most volatile species on Triton and Pluto (methane, CH$_4$, is a distant third), and so dominate seasonal volatile transport across their surfaces. It is entirely reasonable to expect CO and N$_2$ molecules to mix throughout the ice. CO and N$_2$ molecules have similar sizes, shapes, and masses. They have similar volatility, and they are unusual in that they are fully miscible in one another, in both liquid and solid phases. On the other hand, it is entirely possible that the surfaces of Triton and Pluto have regions of largely pure CO ice and largely pure N$_2$ ice. Specifically, N$_2$ is somewhat more volatile than CO, so the two species could separate through a solid state distillation process 
\medskip

Observations of Triton and Pluto lead toward the mixing of CO in N$_2$ ice. 
In the case of Triton, \cite{gru10} found similar patterns of longitudinal variation of CO and N$_2$ ice. For Pluto, \cite{gru16} analyzed New Horizons spacecraft data and found CO and N$_2$ ice coexist in Sputnik Planitia, {\it i.e.} the western lobe of the heart-shaped Tombaugh Regio. The similar location of the two species on the two bodies suggest the CO molecules are mixed in the N$_2$ ice. However, as of yet there is no spectroscopic evidence  on either Triton or Pluto. 
\medskip

Finding spectroscopic evidence first requires an understanding of the physical structure of pure N$_2$ ice and  pure CO ice. Both species undergo a solid-solid phase transition between a higher temperature $\beta$-phase with an orientationally disordered hexagonal crystal structure and a lower temperature $\alpha$-phase with an orientationally ordered cubic structure.  In pure N$_2$, this $\alpha-\beta$ transition occurs at 35.61 K \citep{sco76}, while in pure CO it occurs at a much warmer 61.6 K \citep{bm65}.  Since the two are fully miscible, the transition temperature can be expected to vary as a function of composition, and a phase diagram published by \cite{aw66}  from x-ray diffraction shows exactly that.  The binary phase diagram was further explored by \cite{vj07} using infrared spectroscopy, resulting in $\alpha$-$\beta$ phase boundary curves somewhat different from the Angwin and Wasserman curves.  
\medskip

Besides experiments to discern the phase of the ice, additional  experiments are essential to relate the phase to a spectroscopic signature. Only then can the laboratory spectra be compared to spectra of Triton and Pluto.  \cite{qui97} made an extensive laboratory study of CO diluted in N$_2$. Their work focused on the frequency, width, and intensity of the CO first vibrational overtone (0-2) at 4252 cm$^{-1}$ because of its potential application to analyzing spectra of Triton and Pluto. Quirico and Schmitt found that the phase of the N$_2$ ice made a difference in the appearance of the CO band. In particular, the CO (0-2) band was much more intense and narrower in the ordered $\alpha$-phase than in the disordered $\beta$-phase. In addition, they found a difference in the frequency of the (0-2) band in the $\alpha$ vs. $\beta$ ice.  Besides the common isotope of CO, Quirico and Schmitt reported frequency measurements of the (0-2) band for uncommon isotopes of CO in the $\alpha$-phase. Of particular importance to the work we report here, Quirico and Schmitt also reported a new and unidentified band at 4467.3 cm$^{-1}$. See their Figure 1 and Table 1.  
\medskip

Considering the importance of ice phase on the spectroscopic signature of a CO/N$_2$ ice mixture, and the difference between the two published phase diagrams, we decided to revisit the phase diagram, but instead of using transmission spectroscopy, we opted to use Raman spectroscopy.  
Our rationale for use of Raman instead of transmission spectroscopy was that N$_2$ is a symmetric molecule with no dipole, so its infrared spectral absorption is extremely weak, involving a temporary induced dipole moment from interaction with a neighboring molecule \citep{sg66,se71}.  The weakness of the infrared absorption by N$_2$ creates difficulties for detecting its features in transmission spectra of thin or highly scattering samples. CO and N$_2$ both have strong Raman bands \citep{cl69}, so use of Raman spectroscopy to track phase changes in N$_2$/CO mixtures obviates difficulties associated with the weakness of N$_2$ absorption.
\medskip

In addition to the Raman spectroscopy experiments designed to clarify phase boundaries, we describe a series of laboratory transmission spectroscopy experiments designed to identify the 4467.3 cm$^{-1}$ band in the spectra of \cite{qui97}.  Our experiments show the band is due to the simultaneous vibrational excitation of a CO molecule and an adjacent N$_2$ molecule in the ice by a single photon. We term such a band a two-molecule combination band, which is also known in the literature as a "dimol absorption band," dimol meaning "two-molecule." For example, see \cite{ida10} and \cite{taj17}.  Previous identifications of two-molecule combination bands in spectra of ice samples include single photons exciting the fundamental mode of adjacent N$_2$ molecules \citep{gru93} and adjacent N$_2$ and O$_2$ molecules \citep{min06} as well as single photons exciting an electronic transition in adjacent O$_2$ molecules \citep{lan62,spen02}.

\medskip

The CO-N$_2$ two-molecule combination band  has the potential to further inform us about the state of ices on icy outer solar system bodies.  In particular,  besides our laboratory work,  we present a near-infrared spectrum of Triton from the Gemini 8-meter telescope in Chile and the Immersion Grating Infrared Spectrograph (IGRINS). The spectrum shows the CO-N$_2$ combination band. The detection conclusively shows that a significant amount of CO is indeed diluted in N$_2$ ice on the surface of Triton. 
\medskip

\section{Overtone and Combination Bands}

Since the laboratory transmission spectra we describe below contain overtone bands and a newly identified two-molecule combination band, it's important to elucidate the difference between overtone and combination bands. 
\medskip


\begin{figure}[t!]
\epsscale{1.2}
\plotone{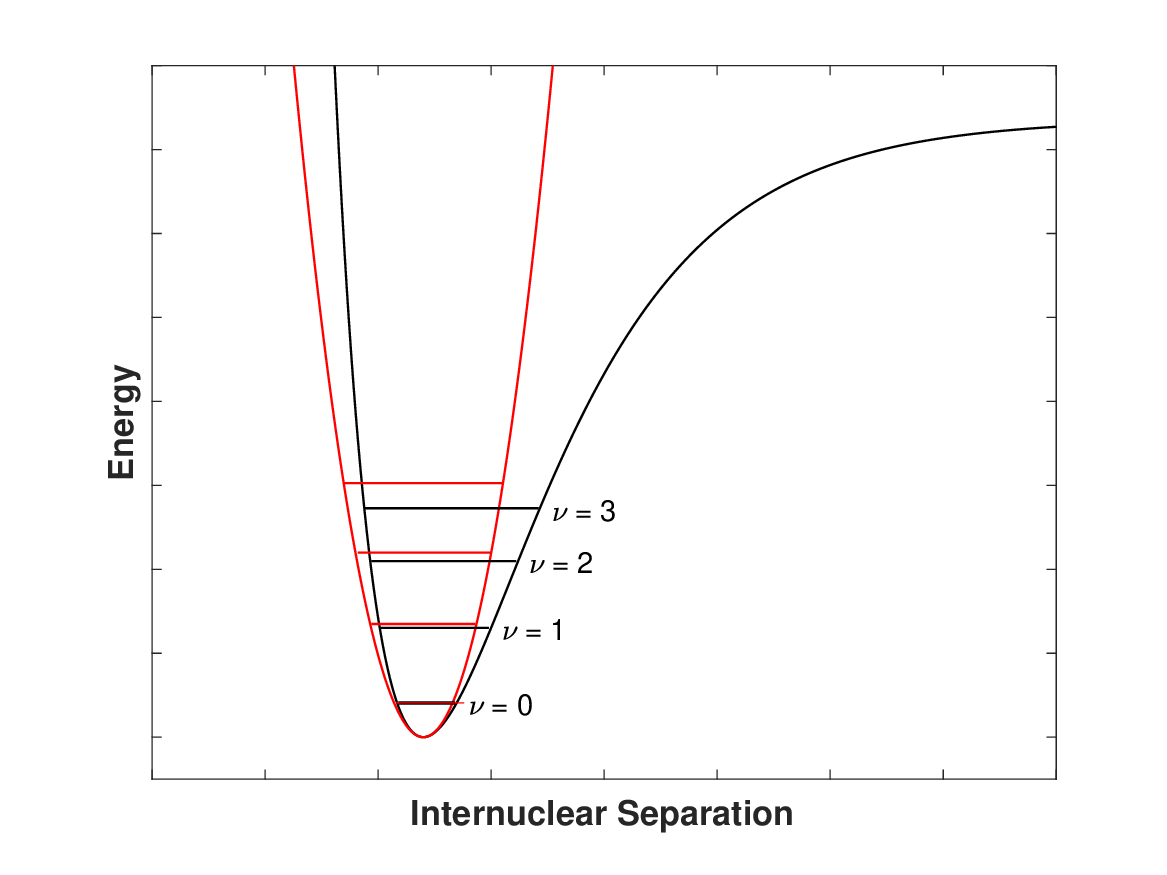}
\caption{The harmonic oscillator potential (red line) and more realistic Morse potential (black line) for vibration of a generic diatomic molecule. The energy levels of the Morse potential are not evenly spaced but decrease in spacing as $\nu$ increases. Of importance here, the frequency of a photon required to excite the molecule from $\nu$ $=$ 0 to $\nu$ $=$ 2, {\it i.e.} the first overtone (0-2),  is not  twice the frequency of a photon required to excite the molecule from $\nu$ $=$ 0 to $\nu$ $=$ 1, {\it i.e.} the fundamental (0-1). Rather, the frequency of the first overtone is less than twice the frequency of the fundamental. \label{fig:f1}}
\end{figure}


In Figure \ref{fig:f1}, we show a harmonic  potential approximation to vibration of a generic diatomic molecule (red line). It gives a good approximation of the fundamental excitation of a vibrational mode (0-1). However, the equally spaced energy levels and a selection rule that requires $\Delta{\nu}$ $=$ $\pm$ 1 predicts a spectrum of only one band whereas we know spectra of real diatomic molecules exhibit more than one band. 
\medskip

An anharmonic  potential is a much better approximation to vibration of a diatomic molecule. In Figure \ref{fig:f1}, we show the Morse potential (black line) which is an example of an anharmonic potential commonly used to describe bond vibration. Notice that unlike the harmonic oscillator potential, the energy levels are not evenly spaced but decrease in spacing as $\nu$ increases. In addition, the anharmonicity breaks the selection rule $\Delta\nu$ $=$ $\pm$1. So, it is possible for photons to excite the molecule from $\nu$ $=$ 0 to $\nu$ $=$ 2 (first overtone) and $\nu$ $=$ 0 to $\nu$ $=$ 3 (second overtone) and so on. An important point to draw from Figure \ref{fig:f1} for our work here is that the first overtone is not excited by a photon with exactly twice the frequency of the fundamental. Rather, it is excited by a photon with a frequency less than twice the frequency of the fundamental.  The same is true of the second overtone. It is not excited by a photon with a frequency three times the fundamental. Rather it is excited by a photon with a frequency less than three times the fundamental.  
\medskip

Combination bands commonly occur when a photon simultaneously excites two or more fundamentals of a single molecule with three or more atoms, {\it i.e.} molecules with more than one vibration mode. For example, a single photon could simultaneously excite the $\nu_1$ O$-$H symmetric stretch fundamental of a H$_2$O molecule at 3657 cm$^{-1}$ and the $\nu_2$ bend fundamental of the same H$_2$O molecule at 1595 cm$^{-1}$  to create a $\nu_1$ $+$ $\nu_2$ combination band at about 5252 cm$^{-1}$. 
\medskip

Here, we report combination bands that result from photons that simultaneously excite the fundamentals of two adjacent molecules, N$_2$ and CO,  rather than  photons that excite two fundamentals in a single molecule. 
\medskip

\section{Experimental Set Up}


\begin{figure*}[t!]
\gridline{\fig{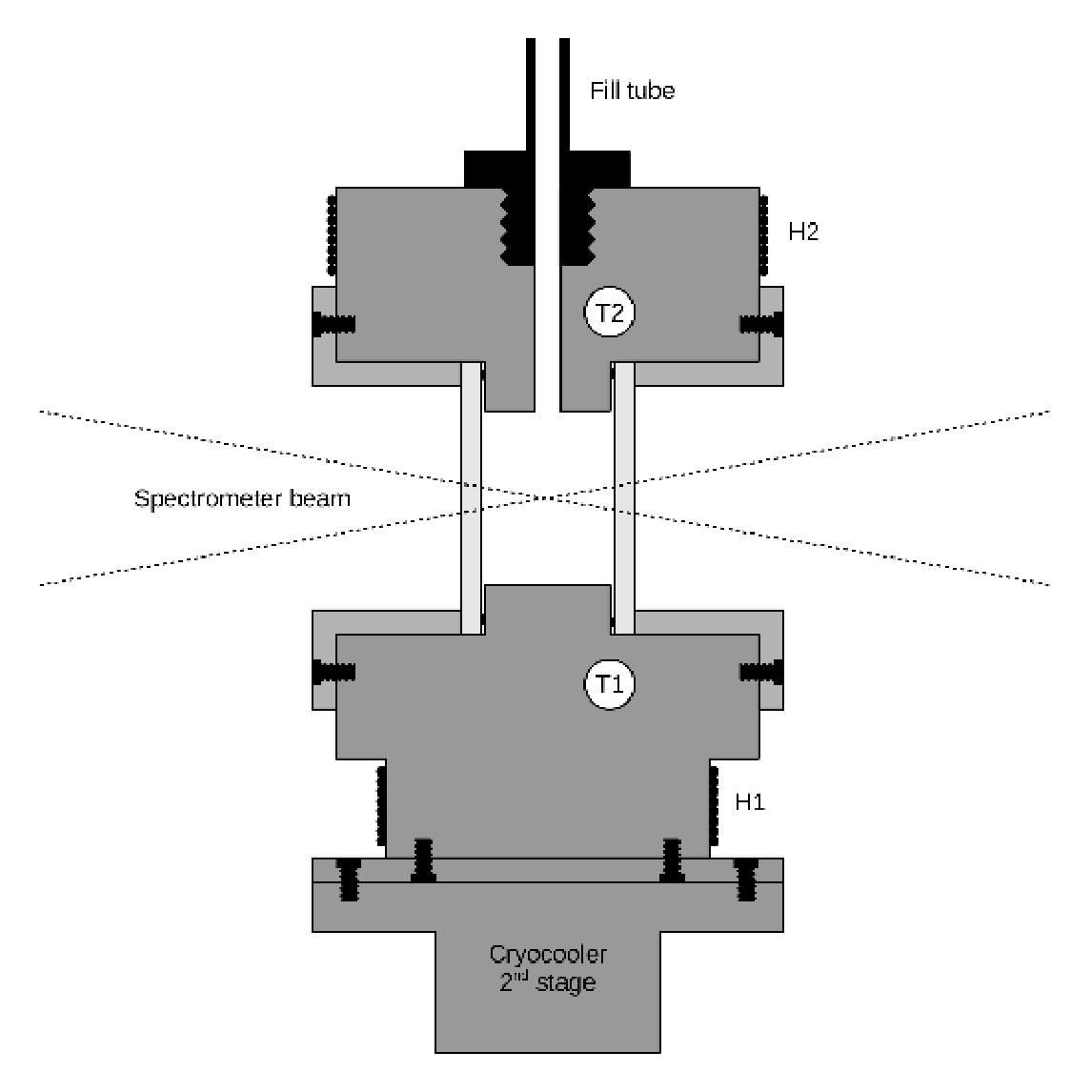}{0.4\textwidth}{(a)}
          \fig{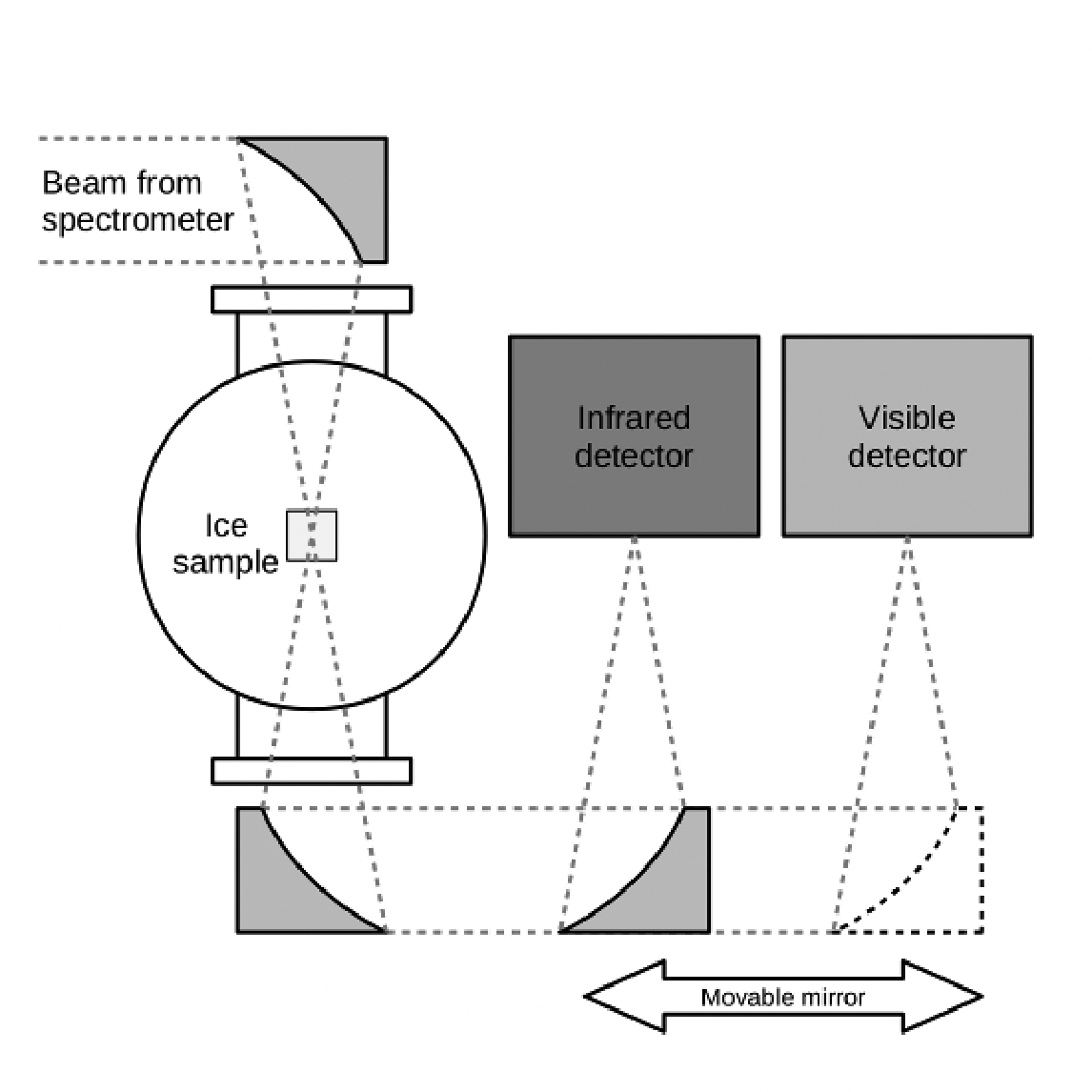}{0.4\textwidth}{(b)}
          }
\caption{(a) Schematic diagram of the cell in cross section as seen from the side. The cell is mounted on top of the cryocooler.  Gas enters the cell from above via a fill tube. The dotted lines represent the spectrometer beam through the sample. Thermometers T1 and T2 and heating elements H1 and H2 control the temperature of the sample. Further details concerning the cell are described in \cite{teg10} and \cite{gru11}. (b) The optical train in our experiment as seen from above. The spectrometer beam is represented by dashed lines. The first  off-axis paraboloid mirror focuses the beam into the sample volume, then two more identical mirrors re-focus the beam onto one of two detectors, an infrared or visible detector. The third mirror in the optical train is movable and enables us to select the infrared or visible detector. Only the infrared detector was used in the experiments described here. \label{fig:f2}}
\end{figure*}


The experiments reported here were carried out in the Astrophysical Materials Laboratory located in the Department of Physics and Astronomy of Northern Arizona University. Ice samples were crystallized in a cell fitted with windows to allow a spectrometer beam to pass through the ice (Figure \ref{fig:f2}a). A detailed description of the facility was published in \cite{teg10}. Subsequent to that paper, we made several important  improvements to the facility including the addition of a mercury cadmium telluride (MCT) type-A 
detector cooled with liquid nitrogen, off-axis aluminum paraboloid mirrors (Figure \ref{fig:f2}b), and a 2 L mixing volume. Detailed descriptions of these improvements were published in \cite{gru11}.
\medskip

Samples were grown as follows. High pressure cylinders of CO (purity $>$ 99.99 $\%$) and N$_2$ (purity  $>$  99.9 $\%$ ) supplied by Airgas were connected through separate pressure regulators and valves to the mixing volume. We mixed gasses at room temperature in the mixing volume. Then, we opened a valve which allowed the gas to flow into the empty, cold cell, condensing it as a liquid. We froze the liquid by reducing the temperature in the cell at a rate of 0.1 K per minute. During the freezing process, we maintained a vertical thermal gradient of about 2 K across the sample by means of heaters (Figure \ref{fig:f2}a). In this way, the sample froze from the bottom upward, with the location of the freezing front controlled by the cell temperature. Once the sample was frozen, we removed the vertical thermal gradient. 
\medskip

Spectra were recorded with a Nicolet iS50 Fourier transform infrared (FTIR) spectrometer at a sampling interval of 0.24 cm$^{-1}$, resulting in a spectral resolution of 0.6 cm$^{-1}$  (full width at half maximum of unresolved lines). We averaged over spectral scans to improve the signal/noise ratio. After we recorded our initial ice spectrum, we ramped down to lower temperatures at  0.1 K per minute. 
\medskip


\begin{figure*}[t!]
\gridline{\fig{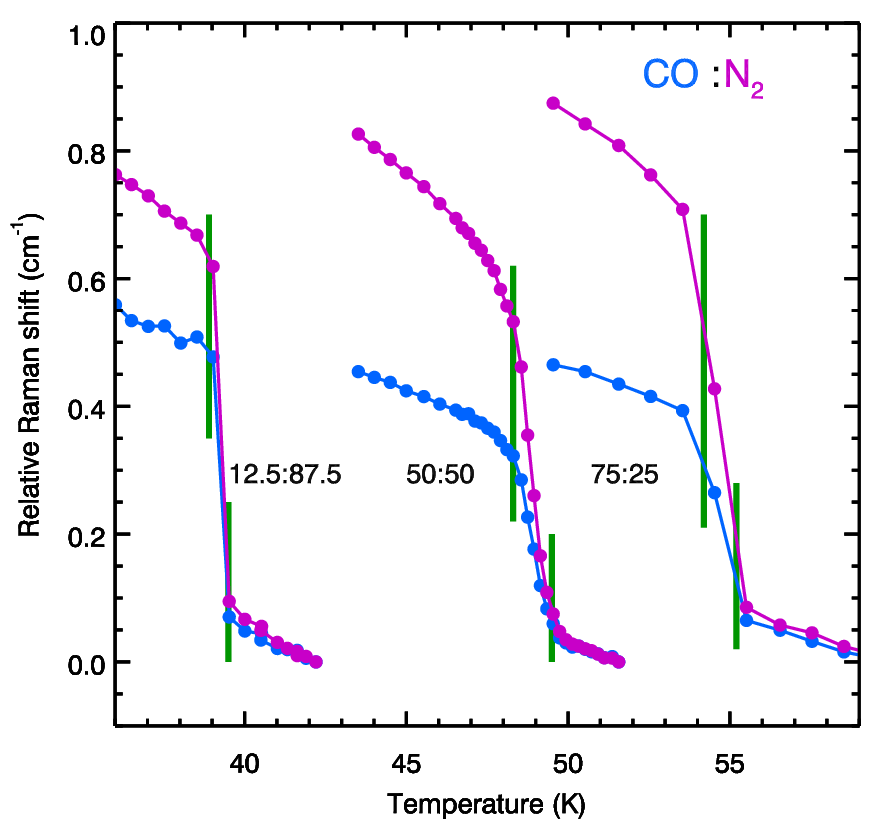}{0.4\textwidth}{(a)}
          \fig{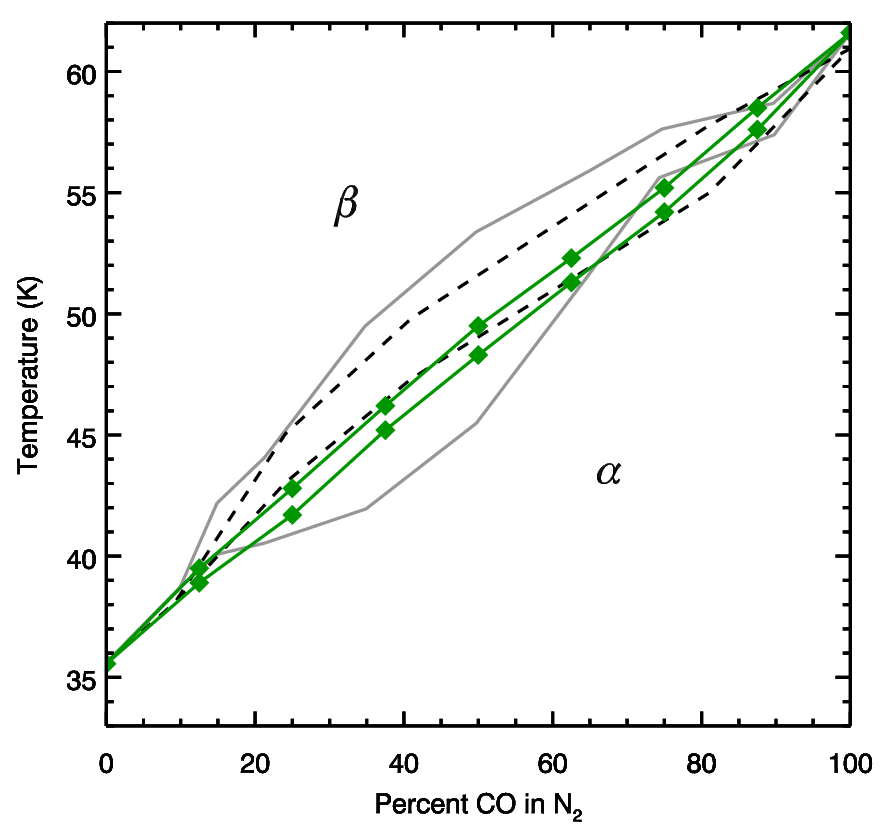}{0.4\textwidth}{(b)}
          }
\caption{(a) Relative Raman shift vs. temperature for CO (blue dots) and N$_2$ (magenta dots) for three different CO/N$_2$ compositions. As temperature gets smaller for each composition and the $\beta$-ice begins to transform into the $\alpha$-ice, the rate of change in relative shift with temperature becomes much greater. The two vertical green lines at higher and lower temperature for each sample mark the start of formation of $\alpha$-ice and the last remnants of $\beta$-ice, respectively. (b) Phase boundaries from our Raman study for CO/N$_2$ ice (green dots and lines) along with the phase boundaries of \cite{aw66} (gray curves) and \cite{vj07} (dashed black curves). Our results show that the two-phase region between the two curves is narrower than found in earlier studies. \label{fig:f3}}
\end{figure*}


Raw transmission spectra were converted to absorption coefficient spectra as follows. First, we removed the instrumental signature and water vapor lines by dividing each raw transmission spectrum by a background spectrum, {\it i.e.} a spectrum taken just before insertion of the sample and/or just after blowing off the sample.  Then, we removed a low-amplitude, high-frequency interference pattern due to the windows in our cell by using a Fourier filter.  Finally, we used the Beer-Lambert  law to compute  absorption coefficient spectra, $\alpha(\nu)$. 
\medskip

Using the same sample preparation procedures described above, we probed
our samples using a Kaiser Optical Systems Rxn1 Raman instrument with a
400 mW 785 nm laser.  The laser beam traveled through an optical fiber and,
after emerging from the fiber, was focused just inside the window of the
sample cell by means a 5.5 inch focal length lens.  Scattered photons
were focused by the same lens back into the optical fiber, which returned
them to a grating spectrometer that recorded wavelengths from about
791 to 1074 nm, corresponding to wavenumbers from 100 to 3425 cm$^{-1}$ below
that of the laser.  These are photons that have lost energy to the
material, a process known as Stokes Raman scattering (as opposed
to anti-Stokes scattering in which photons  gain energy from the
material).
\medskip

\section{Experimental Results}

\subsection{Raman Spectroscopy}

Figure \ref{fig:f3}a shows how we used Raman spectra to determine the temperatures of phase transitions for three different CO and N$_2$ compositions. In both the $\alpha$-  and $\beta$-phases of each of the three samples, the Raman shift slowly increased 
with decreasing temperature.  Furthermore, the shifts were much higher in $\alpha$-phase than in $\beta$ phase.   Of particular importance here, during the formation of the first $\alpha$-ice and until the disappearance of the last $\beta$-ice, the rate
of change in Raman shift with temperature was much faster than when the ice was all $\beta$-ice or all $\alpha$-ice.  The location in Figure \ref{fig:f3}a where the slope
changed (green vertical lines) reveal the upper and lower temperature limits of the two-phase region.  In Figure \ref{fig:f3}b, we plot the upper and lower temperature limits of the two phase region for the three compositions in Figure \ref{fig:f3}a along with four other compositions. In addition, we plot the phase boundaries from \cite{aw66} (gray curves) and from \cite{vj07} (dashed black curves). Our results show that the two-phase region between the two curves is narrower than found in earlier studies. Figure \ref{fig:f3}b gives us confidence we know the phase of our ice samples in our transmission spectroscopy experiments. 
\bigskip

\bigskip

\subsection{Transmission Spectroscopy}

Below we describe the results of our transmission spectroscopy experiments designed to identify the 4467.3 cm$^{-1}$ band reported by \cite{qui97}. 
\bigskip

\subsubsection{$\alpha$-Ice vs. $\beta$-Ice}

\begin{figure}[t!]
\epsscale{1.2}
\plotone{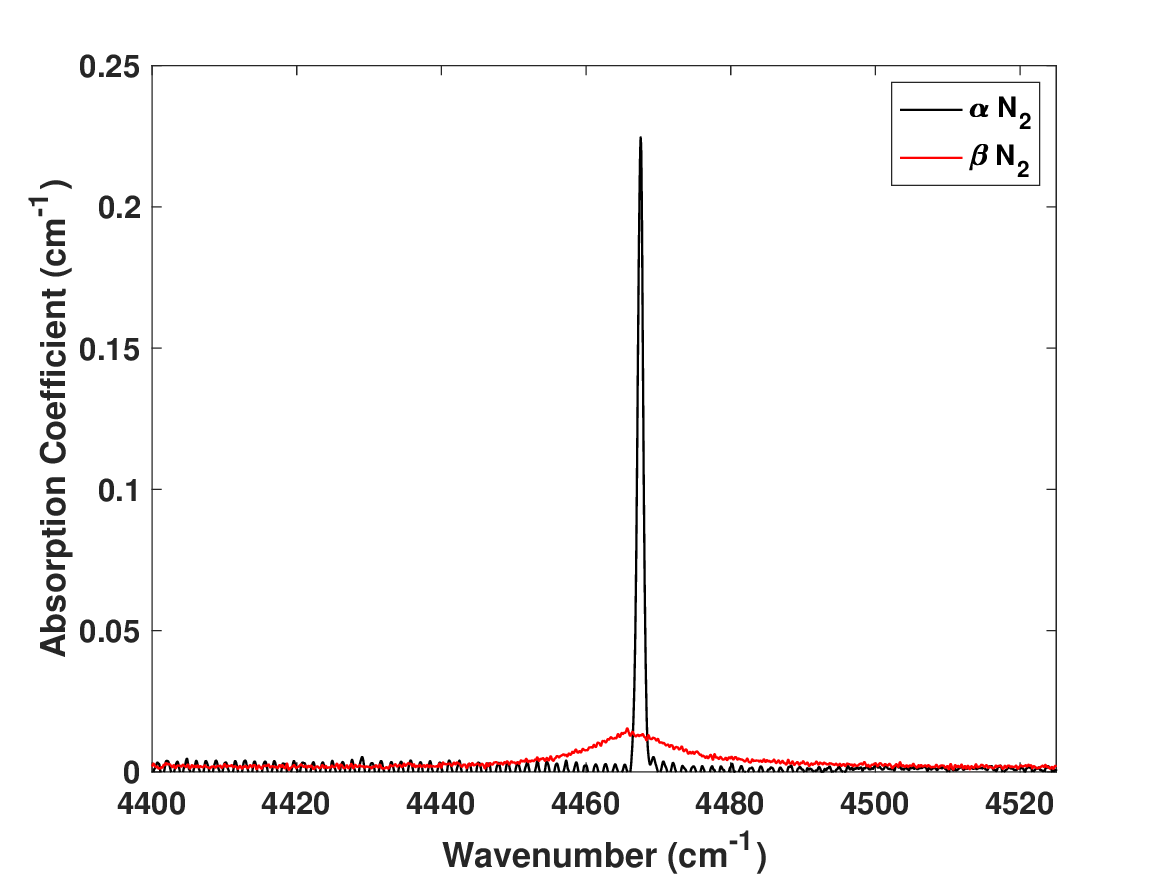}
\caption{The unidentified band in our spectrum of a sample of 4\% CO and 96\% N$_2$ ice at T $=$ 33 K ($\alpha$-phase, black line) and T $=$ 60 K ($\beta$-phase, red line). The band in the $\alpha$-phase is much more intense and narrower than the band in the $\beta$-phase. Peak absorption occurs at 4467.5 cm$^{-1}$ in the $\alpha$-phase and 4466.5 cm$^{-1}$ in the $\beta$-phase.\label{fig:f4}}
\end{figure}

In Figure \ref{fig:f4}, we plot spectra of a 4\% CO and 96\% N$_2$ ice sample at temperatures of 60 K ($\beta$-phase, red line) and 33 K ($\alpha$-phase, black line). From Figure \ref{fig:f3}b, it is clear that the sample was exclusively $\beta$-ice at 60 K and exclusively $\alpha$-ice at 33 K.  In other words, neither spectrum was sampling both phases.  As expected, the unidentified band is more intense and narrower in the $\alpha$-phase than in the $\beta$-phase. It is also clear from Figure \ref{fig:f4} that the peak absorption of the band occurs at different frequencies in $\alpha$-ice and $\beta$-ice. In particular, we measured a frequency of 4467.5 cm$^{-1}$ in the $\alpha$-phase and 4466.5 cm$^{-1}$ in the $\beta$-phase. We point out that our frequency and intensity measurements for the band are in good agreement with  \cite{qui97}.  Specifically, we measured a frequency and intensity of 4467.5 cm$^{-1}$ and  0.22 cm$^{-1}$ and \cite{qui97} measured a frequency and intensity of 4467.3 cm$^{-1}$ and $\sim$0.3 cm$^{-1}$. Our measurements and those of Quirico and Schmitt were made at nearly identical temperatures and compositions but in different laboratories.
\medskip
\medskip


\begin{figure}[t!]
\epsscale{1.2}
\plotone{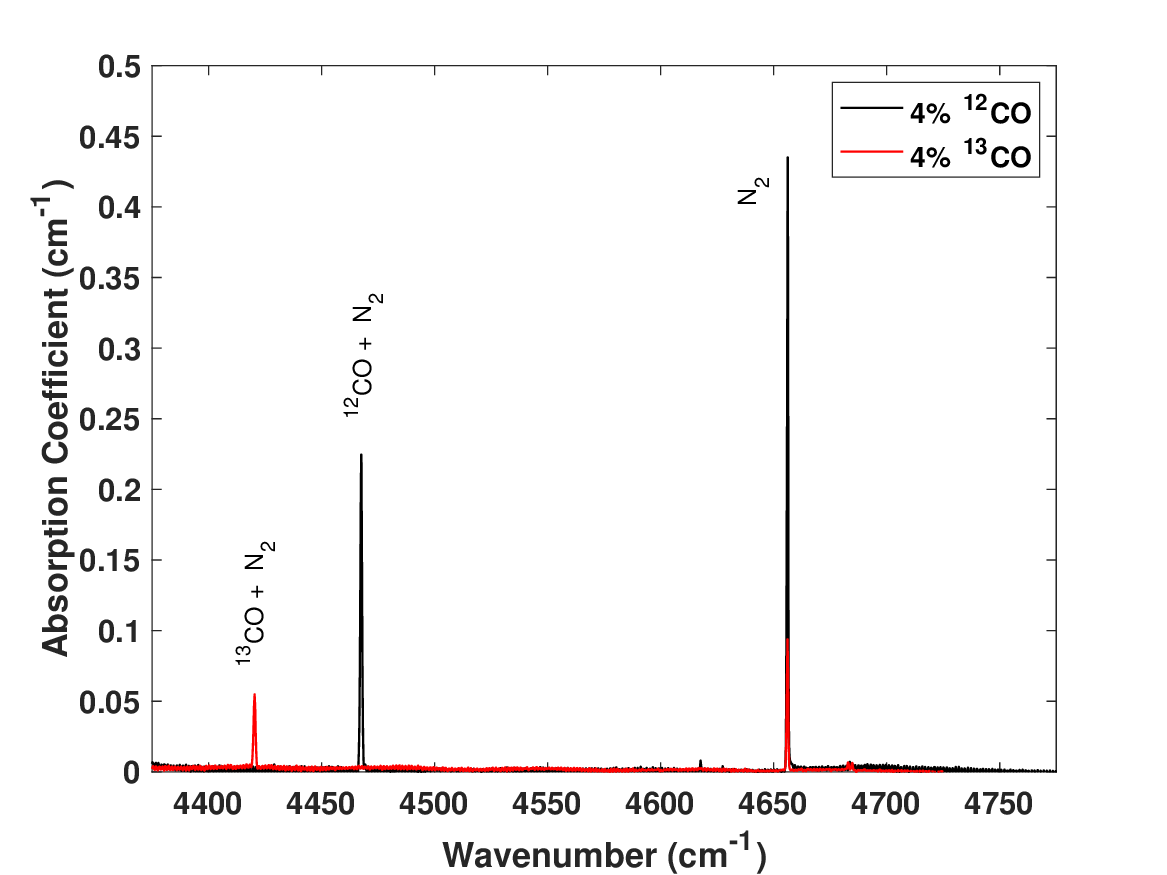}
\caption{The unidentified band and an N$_2$ band in the spectrum of a sample of 4\% $^{12}$C$^{16}$O and 96\% $^{14}$N$_2$ ice at T $=$ 33K (black line). The unidentified band and an N$_2$ band in the spectrum of a sample of 4\% $^{13}$C$^{16}$O and 96\% $^{14}$N$_2$ ice at T $=$ 33K (red line). The peak frequency of the unidentified band shifts from  4467.5 cm$^{-1}$ in the sample with $^{12}$C$^{16}$O to 4420.4 cm$^{-1}$ in the sample with $^{13}$C$^{16}$O.  An examination of the frequencies indicates that the unidentified band arises from single photons exciting adjacent pairs of CO and N$_2$ molecules. We refer to such a band as a two-molecule combination band. \label{fig:f5}}
\end{figure}



\begin{deluxetable}{rccc}[ht!]
\tablewidth{0pt}
\tablecaption{Measured Band Frequencies in $\alpha$-Phase}
\tablehead{
\colhead{Sample} & 
\colhead{Band} &
\colhead{Frequency} &
\colhead{Reference}
\\
\colhead{} & 
\colhead{} &
\colhead{(cm$^{-1}$)} &
\colhead{}}
\startdata
N$_2$ & 0-1 & 2328.1 & a \\
$^{12}$C$^{16}$O & 0-1 & 2139.5 & b\\
$^{13}$C$^{16}$O & 0-1 & 2092.3 & b\\
$^{12}$C$^{18}$O & 0-1 & 2088.4 & b\\
& & & \\
$ ^{14}$N$_2$  $+$ $^{14}$N$_2$ & Combination & 4656.3 & c\\
$^{12}$C$^{16}$O$+$$^{14}$N$_2$ & Combination & 4467.5 & c\\
$^{13}$C$^{16}$O$+$$^{14}$N$_2$ & Combination & 4420.4 & c\\
$^{12}$C$^{18}$O$+$$^{14}$N$_2$ & Combination & 4416.4 & c\\
\enddata
\tablenotetext{a}{\cite{gru93}} 
\tablenotetext{b}{\cite{qui97}} 
\tablenotetext{c}{This work}
\end{deluxetable}

\subsubsection{Frequency Measurements}  

If the unidentified band is due to the simultaneous excitation of a CO and N$_2$ molecule, its frequency should be the sum of a CO frequency and a N$_2$ frequency. The CO fundamental (0-1) has a frequency of  2139.5 cm$^{-1}$  \citep{qui97} and the N$_2$ fundamental (0-1)  has a frequency of 2328.1 cm$^{-1}$ \citep{gru93}, both in the $\alpha$-phase. The sum of these two fundamental frequencies is 4467.6 cm$^{-1}$. As described above, we measured a frequency of 4467.5 cm$^{-1}$ for the unidentified band in the $\alpha$-phase.  As the uncertainty in our frequency measurements is about 0.1  cm$^{-1}$, these frequency measurements are consistent with the unidentified band resulting from the simultaneous absorption of single photons by adjacent pairs of CO and N$_2$ molecules.  
\medskip


\begin{figure*}[t!]
\gridline{\fig{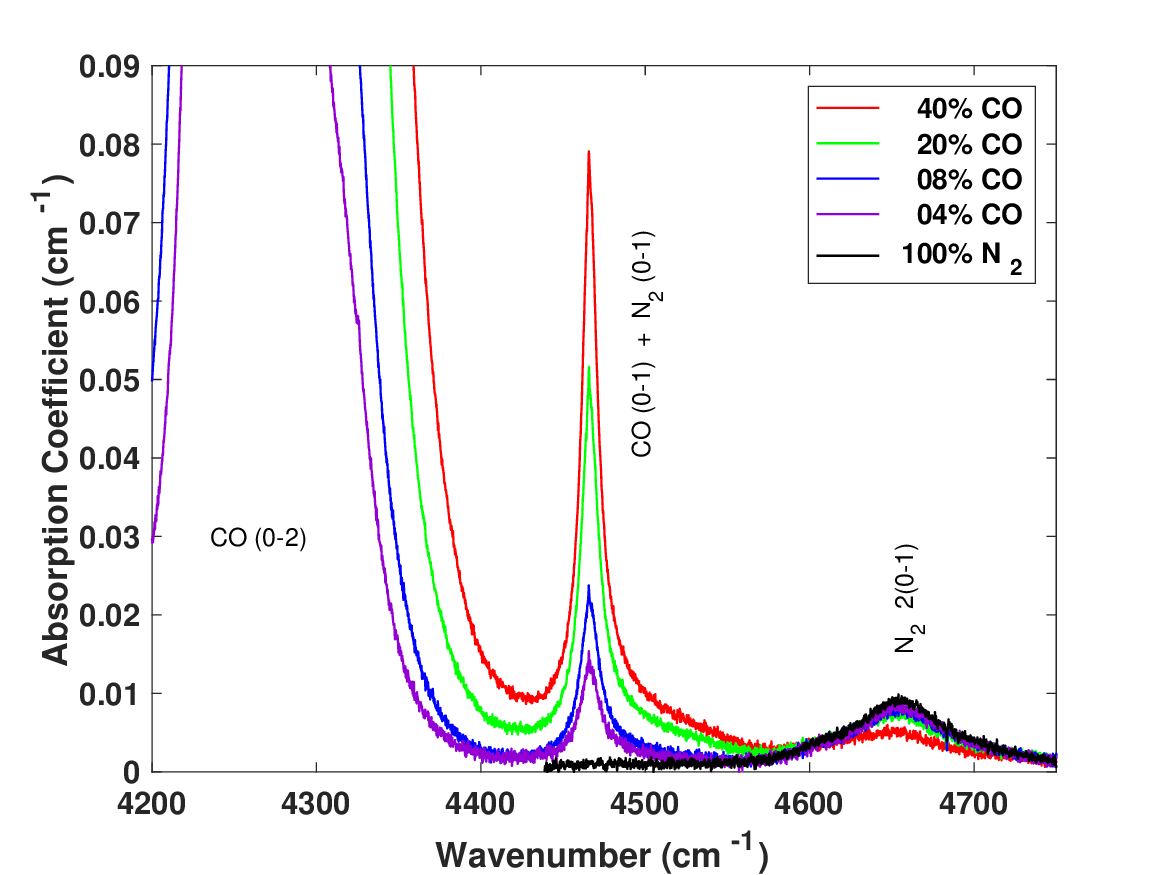}{0.5\textwidth}{(a)}
          \fig{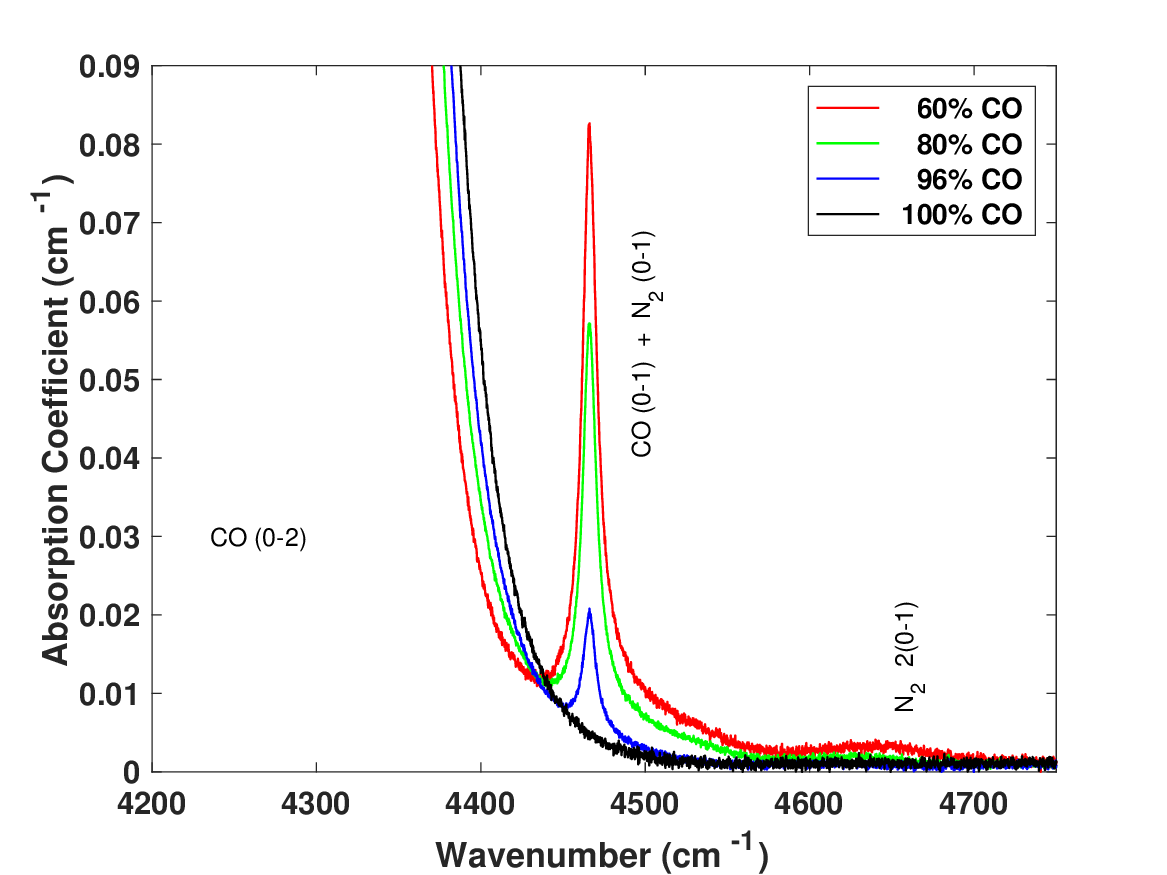}{0.5\textwidth}{(b)}
          }
\caption{(a) Spectra of CO/N$_2$ ice samples with the CO abundance ranging from 0\% to 40\% and all taken at T $=$ 60 K. The N$_2$ ice is in the $\beta$-phase. The spectra show the new band near 4467 cm$^{-1}$.  The new band is not present in the pure N$_2$ sample (black line) and increases in strength with increasing CO abundance. The saturated band at $\sim$4252 cm$^{-1}$ is CO (0-2) and the weak, broad band at $\sim$ 4654 cm$^{-1}$ is N$_2$. (b) Spectra of CO/N$_2$ ice samples with the CO abundance ranging from 60\% to 100\% and all taken at T $=$ 60 K. The N$_2$-ice is in the $\beta$-phase. The spectra show the new band near 4467 cm$^{-1}$. Its band strength at a CO abundance of 60\% is nearly the same as its strength at 40\% (see panel a) and then decreases in strength with increasing CO abundance. The band is not present in the pure CO ice sample. A maximum strength  for samples with nearly equal amounts of CO and N$_2$ and absence in pure N$_2$ and pure CO reinforces the idea that the band is due to photons exciting adjacent pairs of CO and N$_2$ molecules.  \label{fig:f6}}
\end{figure*}


Ices with uncommon isotopic compositions provide a means to further test the above mechanism. Specifically, the frequency of vibrational absorption by a diatomic molecule depends on the masses of the individual atoms. Therefore, changing the mass of one of the atoms, {\it e.g.}  $^{12}$C to $^{13}$C,  will change the frequency of the CO (0-1) band and so should change the frequency of a CO-N$_2$ combination band.  To test our mechanism, we obtained spectra of a 4\% $^{13}$C$^{16}$O $+$ 96\% $^{14}$N$_2$ sample and compared it to spectra of our 4\% $^{12}$C$^{16}$O $+$ 96\% $^{14}$N$_2$  sample. In Figure \ref{fig:f5}, we show  spectra of our 4\% $^{13}$C$^{16}$O sample (red line)  and our 4\% $^{12}$C$^{16}$O sample (black line). The unidentified band shifts from 4467.5 cm$^{-1}$ in the $^{12}$C$^{16}$O sample to 4420.4 cm$^{-1}$ in the $^{13}$C$^{16}$O sample. Note that as expected the $^{14}$N$_2$ band at 4656.2 cm$^{-1}$ did not shift between the two samples.  \cite{qui97} measured a frequency of 2092.3 cm$^{-1}$ for $^{13}$CO (0-1). Combining that frequency with a frequency 2328.1 cm$^{-1}$ for $^{14}$N$_2$ (0-1) we get  4420.4 cm$^{-1}$. In short, there is excellent agreement between  summing the frequencies of $^{13}$C$^{16}$O (0-1) and $^{14}$N$_2$ (0-1) and the frequency of the unidentified band in our spectrum. 
\medskip

We did one more isotope experiment.  Specifically, we obtained spectra of a 4\% $^{12}$C$^{18}$O $+$ 96\% $^{14}$N$_2$ sample and compared it to spectra of our 4\% $^{12}$C$^{16}$O $+$ 96\% $^{14}$N$_2$  sample.  The unidentified band shifted from 4467.5 cm$^{-1}$ in the $^{12}$C$^{16}$O sample to 4416.4 cm$^{-1}$ in the $^{12}$C$^{18}$O sample. \cite{qui97} measured a frequency of 2088.4 cm$^{-1}$ for $^{12}$C$^{18}$O (0-1). Combining that frequency with the $^{14}$N$_2$ (0-1) frequency, we get 4416.5 cm$^{-1}$. In other words, a difference of only 0.1 cm$^{-1}$ between the sum of the fundamental frequencies and our measured frequency for the unidentified band.  Again, the result is consistent with our mechanism of photons exciting adjacent pairs of CO and N$_2$ molecules.  In Table 1, we summarize the measured band frequencies.  

\subsubsection{Stoichiometry}

Yet another way to test our idea that the new band is the result of photons exciting adjacent pairs of CO and N$_2$ molecules is through stoichometry. If our mechanism is correct, the new band should be strongest in spectra of ice samples  where there is a large number of CO molecules adjacent to N$_2$ molecules. In other words, the band should be strongest in ice samples with nearly equal amounts of CO and N$_2$ and absent in spectra of pure CO ice and pure N$_2$ ice. 
\medskip

In Figure \ref{fig:f6}, we show spectra of ten CO/N$_2$ ice samples with increasing CO abundances ranging from no CO, {\it i.e.} pure N$_2$ ice, to pure CO ice. These spectra include the (0$-$2) CO band at 4252 cm$^{-1}$, the new band, and a N$_2$ band at 4654 cm$^{-1}$ .   All spectra were taken at 60 K, so the ice was in the $\beta$ phase.  In Figure \ref{fig:f6}a, we see the band is not present in the spectrum of pure N$_2$ (black line). The band increases in strength as the CO abundance increases in the samples. By comparing Figure \ref{fig:f6}a and Figure \ref{fig:f6}b, we can see the band reaches a maximum strength for samples with nearly equal abundances of N$_2$ and CO, i.e. CO abundances of 40\% and 60\%. In Figure \ref{fig:f6}b, we can see the  band steadily decreases in strength as the CO abundance continues to increase in the samples. The band is not present in the spectrum of a pure CO ice sample (black line). As expected, the new band is strongest in spectra of samples with near equal amounts of CO and N$_2$ and absent in spectra of pure CO and pure N$_2$.  Note that the behavior of the new band is contrary to the behavior of the CO and N$_2$ bands in these 10 spectra which show an increase in their strength with an increase in their abundance. 
\bigskip

\subsubsection{Other Two-Molecule Combination Bands}
The identification of a CO-N$_2$ combination band prompted us to look for similar bands in other samples. Specifically, we looked for evidence of two-molecule combination bands in a pure N$_2$ ice sample and a pure CO ice sample. 
\medskip


\begin{figure}[t!]
\epsscale{1.2}
\plotone{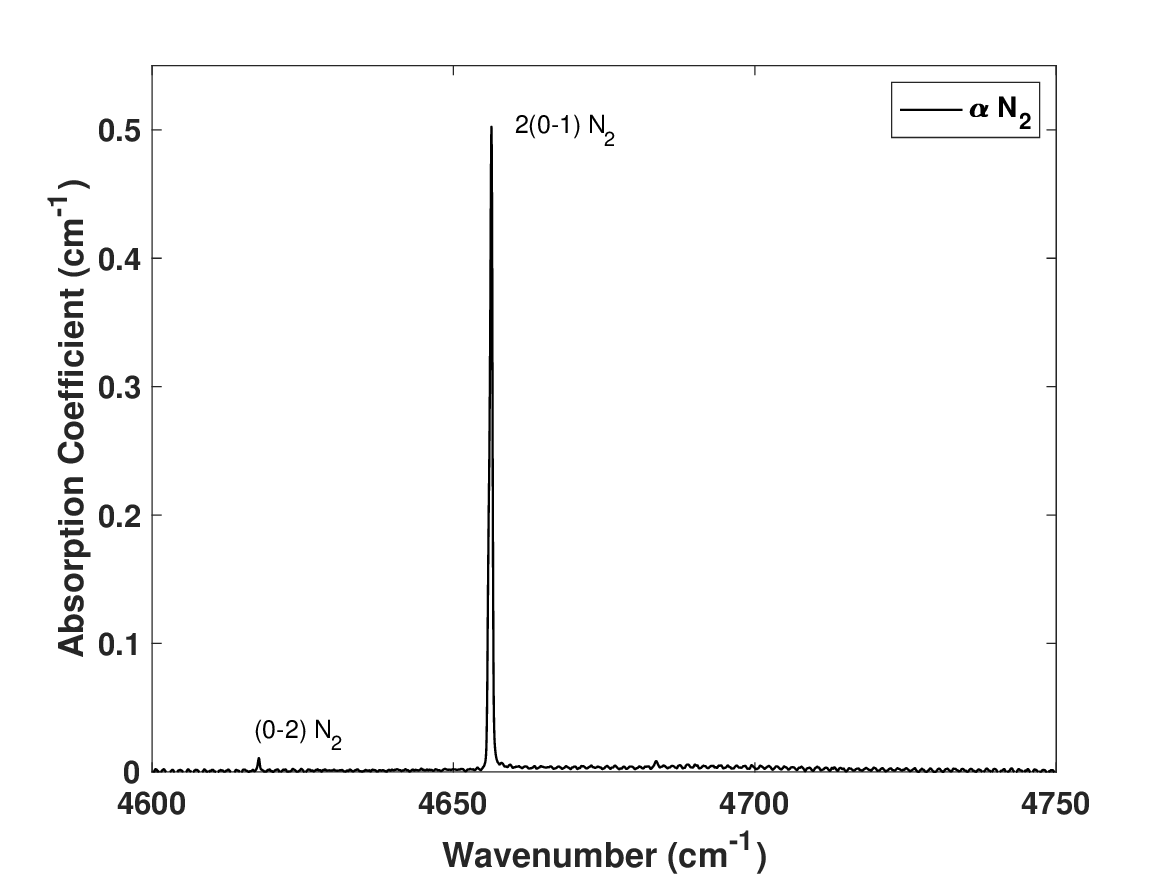}
\caption{Spectrum of pure N$_2$ ice at 32 K ($\alpha$-phase). The strong band has a frequency of 4656.3 cm$^{-1}$, two times the N$_2$ fundamental frequency of 2328.1 cm$^{-1}$. The band is likely the result of single photons vibrationally exciting adjacent pairs of N$_2$ molecules,  {\it i.e.} a two-molecule combination band. We designate it N$_2$ 2(0-1). Presumably, the weak band at 4617.7 cm$^{-1}$ is the first overtone (0-2)\label{fig:f7}}
\end{figure}
\medskip



\begin{figure}[t!]
\epsscale{1.2}
\plotone{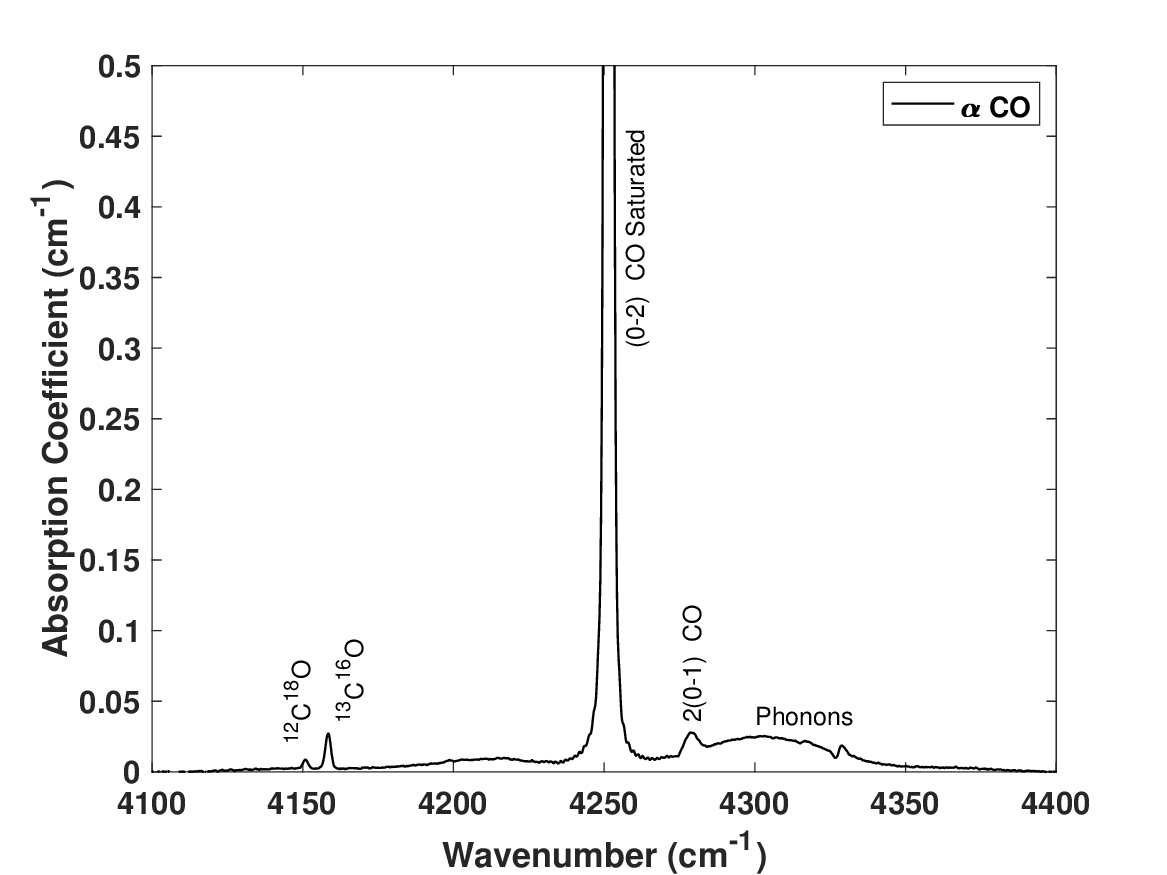}
\caption{Spectrum of pure CO ice at 55 K ($\alpha$-phase). The frequency of the weak band at 4278.9 cm$^{-1}$  has two times the CO  fundamental frequency of 2139.5 cm$^{-1}$. The band is likely the result of single photons vibrationally exciting adjacent pairs of CO molecules. We designate it CO 2(0-1). The saturated band at 4252.2 cm$^{-1}$ is the first overtone (0-2). \label{fig:f8}}
\end{figure}
\medskip


In Figure \ref{fig:f7}, we plot  a spectrum of a pure N$_2$ ice at 32 K. From Figure \ref{fig:f3}b, we see the ice is in the $\alpha$-phase. The strong band in the figure has a frequency of 4656.3 cm$^{-1}$, {\it i.e.} almost exactly two times the N$_2$ fundamental frequency of 2328.1 cm$^{-1}$ . Therefore, the 4656.3 cm$^{-1}$ band is likely a two-molecule combination band. We designate the band N$_2$ 2(0-1). Remember from Figure \ref{fig:f1}, the first overtone of a diatomic molecule occurs at a frequency of less than twice the frequency of the fundamental. Hence, the weaker band at 4617.7 cm$^{-1}$ is likely the first overtone N$_2$ (0-2). It is interesting to note that the planetary science literature designates the N$_2$ band at 4656.3 cm$^{-1}$ as the first overtone rather than a two-molecule combination band (see \cite{pisc88}, \cite{gru91}, \cite{try93}, and \cite{bur10}).
\medskip

In Figure \ref{fig:f8}, we plot  a spectrum of a pure CO ice sample at 55 K. Again, from Figure \ref{fig:f3}b, we see the ice is in the $\alpha$-phase. Note that the weak  band at a frequency of 4278.9 cm$^{-1}$ in Figure \ref{fig:f8} is almost exactly two times the CO fundamental frequency of 2139.5 cm$^{-1}$. Therefore, the 4278.9 cm$^{-1}$ band is likely a two-molecule combination band. We designate it as CO 2(0-1). As expected, the first overtone CO (0-2) occurs at a frequency of 4252.2 cm$^{-1}$, {\it i.e.} less than twice the frequency of the fundamental. Note that the first overtone band (0-2)  is saturated. 
\medskip

\section{Astronomical Spectroscopy}
\medskip
We obtained near-infrared spectra of Triton on the night of 2018 July 2 UT using the Gemini-South 8.1 meter telescope and IGRINS \citep{park14,mace18}.   IGRINS' slit size on the sky was 0.34 arc sec wide by 5 arc sec long. IGRINS provided wavelength coverage from 1.45 $\mu$m  to 2.5 $\mu$m, a resolution ($\lambda$/fwhm) of 45,000, and a sampling of 0.000017 $\mu$m pixel$^{-1}$ at 2.239 $\mu$m (4466.5 cm$^{-1}$).  There were cirrus clouds and the seeing was $\sim$ 0.9 arcsec at the time of observations. Triton was at an airmass of 1.26  and 1.10  at the start and end of observations, respectively.  The observing sequence involved nodding Triton along the slit in an A-B-B-A pattern. The spectra were then reduced using the standard IGRINS Pipeline package \citep{lg17}. The pipeline first removed pixel to pixel variations through flat fielding, it then performed spatial rectification of the two-dimension spectra using flat-lamp echellograms. A wavelength correction was then applied using sky emission and absorption features. Finally, the spectra were optimally extracted from the two dimensional image \citep{hor86}. The output result for each observation of an A-B pairs was a one-dimensional spectrum of wavelength vs. flux for each order. Spectra of an A0V comparison star were taken in order to remove instrumental signatures and telluric bands in the spectra of Triton.  We summed our individual Triton spectra to obtain a single spectrum with a total exposure time of 80 minutes.  Since our objective was to detect the spectrally broad CO-N$_2$ combination band at 2.239 $\mu$m (4466.5 cm$^{-1}$), we used inverse variance weighting to bin the spectrum into blocks of 64 pixels, and thereby improve the signal-to-noise ratio of the Triton spectrum. The binned spectrum had a resolution of $\lambda$/$\Delta\lambda$ $=$ 2500.
\medskip


\begin{figure}[t!]
\epsscale{1.2}
\plotone{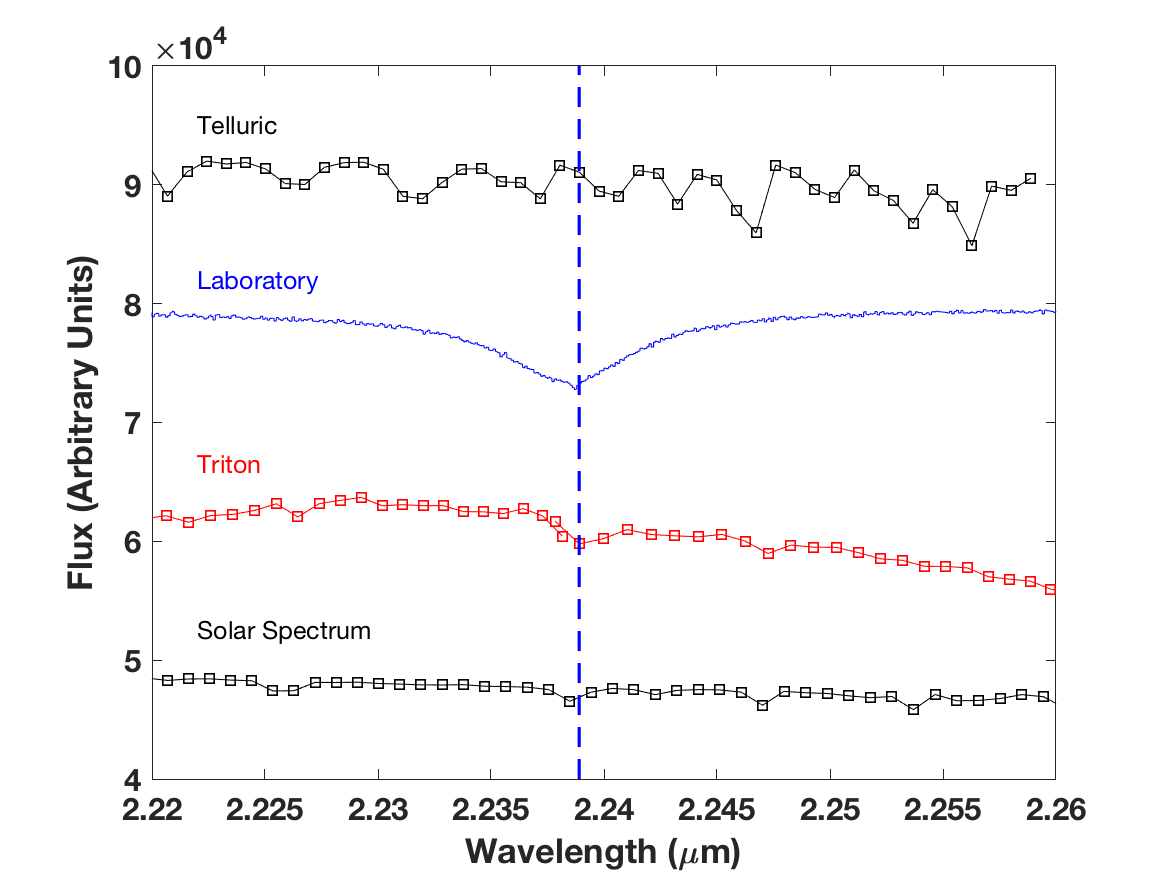}
\caption{A portion of an 80-minute, binned-spectrum of Triton taken with the 8.1-meter Gemini-South telescope and the IGRINS spectrometer (red squares). The broad absorption in the Triton spectrum is consistent with the broad absorption of the two-molecule combination band at 2.239 $\mu$m (4466.5 cm$^{-1}$) band in our laboratory transmission spectrum (blue line). Both broad bands are inconsistent with  the telluric (black squares at top of figure) and the solar (black squares at bottom of figure) spectra.  \label{fig:f9}}
\end{figure}
\medskip


In Figure \ref{fig:f9}, we provide evidence for the detection of the two-molecule combination band at 2.239 $\mu$m (4466.5 cm$^{-1}$) in the spectrum of Triton (red squares). In addition, we show a transmission spectrum of an 8\% CO and 92\% N$_2$ ice sample at 60 K, {\it i.e.} a $\beta$-ice (blue line). We use a 60 K sample because of convenience for the laboratory experiments and not to derive a surface temperature for Triton which is $\sim$ 40 K. The lab band associated with $\beta$-ice at 60 K is nearly identical to the lab band associated with $\beta$-ice at 40 K. For comparison, we show absorption due to Earth's atmosphere (black squares at top of figure) and reflected sunlight, {\it i.e.} Fraunhofer lines (black squares at bottom of figure). The telluric and solar spectra are binned to the same resolution as the binned Triton spectrum,  {\it i.e.} $\lambda$/$\Delta\lambda$ $=$ 2500. The vertical dotted line marks the wavelength of maximum absorption by the broad band in our Triton spectrum. The band in our Triton spectrum coincides with the 2.239 $\mu$m (4466.5 cm$^{-1}$) band in the laboratory spectrum. A comparison of the Triton and telluric spectra shows our A0V star observations did an excellent job of cancelling atmospheric lines.  A comparison of the Triton and solar spectra shows that three weak solar bands contaminate the Triton spectrum, one of which coincides with the two-molecule combination band at 2.239 $\mu$m (4466.5 cm$^{-1}$).  However, it appears the solar band is much weaker and narrower than the two-molecule combination band.  On the basis of these comparisons, we have a high confidence in the detection of the band in the spectrum of Triton. Furthermore, \cite{mer18} detected a broad and weak absorption band at 2.239 $\mu$m in their spectrum of Triton taken with the 8.2-m UT4 telescope in Chile, but they did not identify it.

We caution the reader to not over interpret the data in Figure \ref{fig:f9}.  The Triton spectrum is a reflectance spectrum and the lab spectrum is a transmission spectrum. Hence, it is not possible to quantitatively fit the band in the lab spectrum to the band in the Triton spectrum in Figure \ref{fig:f9}. Such a fit requires us to first convert the laboratory transmission spectrum  to a reflectance spectrum using a radiation transfer model, {\it e.g.} a Hapke model.  Such a conversion is beyond the scope of this paper.

\section{Conclusions}
\medskip

We carried out extensive experiments in order to identify a band at 4467.3 cm$^{-1}$ that was first detected in a sample of CO (4\%) isolated in solid $\alpha$-N$_2$ ice \citep{qui97}. Our laboratory results show that the band is strongest in samples with near equal amounts of CO and $N_2$ and is not present in either pure CO or pure N$_2$ samples. In addition, we found that summing the frequencies of the CO (0-1) fundamental and the N$_2$ (0-1) fundamental agreed with the frequency of the band under study. We performed a similar analysis on two additional samples $-$ one of $^{13}$C$^{16}$O/$^{14}$N and the other of $^{12}$C$^{18}$O/$^{14}$N. In all samples, we found summing the frequencies of the CO and N$_2$ fundamentals agreed with the frequencies of the band under study. These experiments indicate that photons are exciting adjacent CO and N$_2$ molecules to produce the  band. We call the  band a two-molecule combination band. 
\medskip

In addition to our laboratory experiments, we obtained an 80 minute spectrum of Triton using the Gemini South 8-meter telescope and the IGRINS spectrometer. The spectrum shows clear evidence for the two-molecule combination band at 2.239 $\mu$m (4466.5 cm$^{-1}$). \cite{mer18} found the same band in their VLT spectrum of Triton. However, they did not identify the band.  The presence of the band in spectra of Triton indicates that CO and N$_2$ are intimately mixed in the surface ice rather than separated into spatially distinct CO and N$_2$ regions. 
\medskip

The CO-N$_2$ combination band has the potential to better inform us about the ice composition of icy dwarf planets. For example, future ground-based observations could determine if the band strength depends on the Triton longitude.  Perhaps a future spacecraft mission to Pluto will image Sputnik Planitia  in the 2.239 $\mu$m (4467 cm$^{-1}$) band. Such an image would provide a wealth of stoichiometric information about the surface ices and so help constrain geologic and atmospheric processes on Pluto.  
\medskip

We thank the NASA Outer Planets Research program (NNX11AM53G), NASA Solar Systems Workings program (80NSSC19K0556), John and Maureen Hendricks Foundation, Technology Research Initiative Fund at NAU, and National Science Foundation Research Experience for Undergraduates program at NAU (AST-1461200)  for their financial support. 
\medskip

This work was based on observations obtained at the Gemini Observatory, which is operated by the Association of Universities for Research in Astronomy, Inc., under a cooperative agreement with the NSF on behalf of the Gemini partnership: the National Science Foundation (United States), National Research Council (Canada), CONICYT (Chile), Ministerio de Ciencia, Tecnolog\'{i}a e Innovaci\'{o}n Productiva (Argentina), Minist\'{e}rio da Ci\^{e}ncia, Tecnologia e Inova\c{c}\~{a}o (Brazil), and Korea Astronomy and Space Science Institute (Republic of Korea).
\medskip

This work used the Immersion Grating Infrared Spectrometer (IGRINS) that was developed under a collaboration between the University of Texas at Austin and the Korea Astronomy and Space Science Institute (KASI) with the financial support of the US National Science Foundation under grants AST-1229522 and AST-1702267, of the University of Texas at Austin, and of the Korean GMT Project of KASI.
\medskip

\software{IGRINS Pipeline (Lee \& Gullikson 2017)}

\end{document}